\RequirePackage{fix-cm}
\documentclass[onecolumn,epjc3]{svjour3}        
\DeclareUnicodeCharacter{0301}{\'a}
\usepackage[
citestyle=numeric,sorting=none]{biblatex}
\bibliography{main}

\smartqed  
\usepackage{graphicx}
\begin{document}

\title{Dielectric Strength of Noble and Quenched Gases for High Pressure Time Projection Chambers
}


\author{L. Norman\thanksref{e1,addr} 
\and K. Silva\thanksref{addr}
\and B.J.P. Jones\thanksref{e2,addr}
\and A.D. McDonald\thanksref{addr}
\and M. R. Tiscareno\thanksref{addr}
\and K. Woodruff\thanksref{addr}
}
\thankstext{e1}{e-mail: logan.norman@mavs.uta.edu}
\thankstext{e2}{e-mail: ben.jones@uta.edu}


\institute{Department of Physics, University of Texas at Arlington, 108 Science Hall, 502 Yates St, Arlington, TX 76019, United States of America \label{addr}
}

\date{Received: date / Accepted: date}

\maketitle

\begin{abstract}
Dielectric breakdown strength is one of the critical performance metrics for pure gases and gas mixtures used in large, high pressure gas time projection chambers.  In this paper we experimentally study dielectric breakdown strengths of several important time projection chamber working gases and  gas-phase insulators over the pressure range 100~mbar to 10~bar, and gap sizes ranging from 0.1 to 10~mm.  Gases characterized include argon, xenon, CO$_2$, CF$_4$, and mixtures 90-10 argon-CH$_4$, 90-10 argon-CO$_2$ and 99-1 argon-CF$_4$.  We develop a theoretical model for high voltage breakdown based on microphysical simulations that use {\tt PyBoltz} electron swarm Monte Carlo results as input to Townsend- and Meek-like discharge criteria. This model is shown to be highly predictive at high pressure, out-performing traditional Paschen-Townsend and Meek-Raether models significantly.  At lower pressure-times-distance, the Townsend-like model is an excellent description for noble gases whereas the Meek-like model provides a highly accurate prediction for insulating gases.
\keywords{Gaseous detectors \and Charge transport and multiplication in gas}
\end{abstract}

\section{Introduction}

Since their introduction in 1974~\cite{Nygren:1974nfi}, time projection chamber (TPC) detectors have become pervasive in neutrino physics and rare event searches~\cite{MicroBooNE:2016pwy,DUNE:2015lol,NEXT:2012zwy,Aprile:2012zx,EXO-200:2012pdt,LUX:2016ggv,ICARUS:2004wqc}).  The working principle of a TPC~\cite{Gonzalez-Diaz:2017gxo} is that ionization electrons or ions created by a charged particle in a liquid or gaseous medium are drifted to a detection plane  preserving the three dimensional event structure.  At the detection plane, the charges can be detected by induction or collection, or amplified through avalanche gain or electroluminescence, depending on the application. In all cases, to drift charges to the detection plane before the event image is rendered un-reconstructable by diffusion or electron attachment, a drift field in the range 100-1000 V/cm must be applied to the active medium.  For meter to tens-of-meter scale devices, this implies a high voltage be delivered to the detector anode or cathode that is of order tens to hundreds of kilovolts.  This implies high voltage engineering challenges to ensure stable and safe detector operation.

For many neutrino physics or rare event detection applications, maximizing target density is of central importance.  Cryogenic liquid time projection chambers are thus often favored due to the increased density of cryogenic liquids over gas phases of the same materials.  In some situations, the increased precision available from tracking in a lower density medium favors the use of gaseous working media, however.  Examples include identification of the topological two-electron signature~\cite{NEXT:2019gtz} and exquisite energy resolution~\cite{NEXT:2019qbo} for neutrinoless double beta decay searches in xenon gas, as employed by the NEXT program; and precise studies of neutrino interactions in the high-rate environment of the LBNF neutrino beam that will be made by an argon-gas-based detector in the DUNE near detector complex~\cite{martin2017pressurized,DUNE:2015lol,,DUNE:2021tad}. In both of these applications, the relevant gas operating pressures are around 10 bar.

The study of high voltage breakdown of materials, both theoretically and experimentally, is a mature subject.  The theory of dielectric discharge in gases traces its origin to the seminal work of Townsend in 1897~\cite{von1957john}.  Nevertheless, fully trustworthy predictions of the breakdown properties of time projection chamber working media remain largely unavailable.  Several recent experimental works have addressed the high voltage breakdown of liquid argon~\cite{acciarri2014liquid,lockwitz2016study,blatter2014experimental,cantini2017first}  and xenon~\cite{tvrznikova2019direct} for time projection chamber applications.  Experimental study of xenon gas and other pure noble elements for time projection chambers at pressures of up to 10 bar have been presented~\cite{rogers2018high,woodruff2020radio}, and various technical solutions to avoid breakdown or damage to time projection chambers in the event of discharge have also been proposed~\cite{asaadi2014testing,auger2014method}.  There is also a wealth of literature on experimental studies of gaseous insulators under pulsed discharge conditions (for example, for an incomplete list, Refs.~\cite{christophorou1976high,cooke1978nature,hernandez2003pulsed,hornbeck1951microsecond,dahl2012obtaining,hunter1986electron,haefliger2018comparison}).  

In this paper we study the dielectric breakdown voltages of high pressure gases and gas mixtures of immediate interest to neutrino detection at 10 bar.  This work is partly motivated both by the practical needs of the ND-GAr DUNE near detector, which is exploring gas mixtures including Ar-CO$_2$, Ar-CF$_4$, and Ar-CH$_4$ for the active medium, as well as insulating gases CO$_2$ and CF$_4$ for a separate buffer region used as a dedicated gaseous insulator~\cite{martin2017pressurized}.  Other mixtures with much larger hydrocarbon fraction are also being considered for studies of neutrino-hydrogen interactions~\cite{Hamacher-Baumann:2020ogq}.  The work also has a more general motivation, to develop and validate a predictive model of breakdown of high pressure gases for time projection chambers from fundamental microphysical considerations, which would also be of relevance to high pressure gas detectors for neutrinoless double beta decay~\cite{NEXT:2012zwy}. Beyond using pure xenon, which has already been characterized for its high voltage strength at 10 bar~\cite{rogers2018high}, such experiments are also exploring various dopants and additives that may impact high voltage strength~\cite{mcdonald2019electron,rogers2020mitigation,felkai2018helium,cebrian2013micromegas,dafni2009energy,gonzalez2015accurate,azevedo2016homeopathic,fernandes2020low}.  Due to additional costs and complexity of handling and recapturing xenon-based gas mixtures, our primary experimental focus in this work falls upon argon-based mixtures, though a comparative study of pure xenon gas validates the developed theoretical treatment for use with xenon as the base gas also.

This paper is structured as follows.  In Sec.~\ref{Theoretical Modeling}, we discuss the physics of high voltage breakdown in the 0.1-10~bar pressure range.  We review some previous models that have been used to characterize high voltage breakdown thresholds, and introduce our microscopic extension of the Townsend and Meek discharge models based on the {\tt PyBoltz}~\cite{al2020electron} swarm simulation package, used to interpret our experimental data. We  present our experimental methods in Sec.~\ref{experimental apparatus}, including a detailed explanation of the apparatus, data collection method, and systematic uncertainty analysis. Sec.~\ref{results}  presents the experimental results and comparison to theoretical predictions.  We conclude with a discussion of the implications of this work for future gas-phase neutrino detectors in Sec.~\ref{sec:Conclusions}.

\section{Theoretical Models of High Voltage Breakdown}\label{Theoretical Modeling}

For the purposes of this work, we define a high voltage breakdown as a condition when the applied potential between two electrodes is sufficient to cause a discharge that can be self sustaining.  This causes a large current to flow leading to rapid drop in the potential difference~\cite{raizeryurip}. The original Townsend breakdown is at its core a steady-state phenomenon.  Under the original formulation, electron avalanche multiplication in a gas generates positive ions that travel to the cathode where their impact with the cathode surface may liberate further electrons, leading to a self-sustained discharge.  The Townsend breakdown criterion can also be applied to non-steady-state scenarios and photon-feedback driven avalanches. Combination of the Townsend model with an empirical parameterization of the Townsend attachment coefficient leads to a description of high voltage breakdown based on Paschen's law. We review this model in Sec~\ref{townsendPaschen}.

Townsend theory was called into question for breakdown of gases at higher pressures, where it is expected that the breakdown mechanism transitions from the Townsend breakdown to a streamer or leader breakdown~\cite{meek1978electrical,raizeryurip}. Under these breakdown mechanisms, space charge of ions at the tip of an advancing column of charge play a key role.  We briefly describe this theory, originated by Meek and Raether, in Sec. \ref{meek}. The transition from Townsend to streamer breakdown is imperfectly described but is generally understood to occur for air at some pressure-distance above 5 bar-cm~\cite{raizeryurip}.  At the pressure range of our experiments, and given the large variety of gases used, it is unclear a priori whether the Townsend or Meek description of the breakdown criterion would be more appropriate.  However, since both models depend on empirical parameters that must be measured for each gas, it is often the case that in practice either can provide a reasonable fit to experimental data over a limited pressure range. 

This dependence on arbitrary parameters rather compromises the predictivity of the two models above.  One of our motivating goals in this work is to develop and test a fully predictive model of high voltage breakdown for TPC gases in the 1-10 bar pressure regime. In Sec. \ref{pyboltz} we describe two such models, which augment the Townsend and Meek discharge models with microphysically calculated values of the Townsend first ionization and attachment coefficients.  Both models are in excellent agreement at high pressure-times-distance $pd$ and out-perform both the Paschen-Townsend and Meek-Raether models with their conventional parameters when compared to our data, providing a strong description of the breakdown strengths of all the tested gas mixtures over the pressure range 10$^2$-10$^4$ Torr-cm.   At the lowest values of $pd$ the {\tt PyBoltz}-augmented Meek and Townsend predictions diverge and we find that breakdowns in noble gases are well predicted by the Townsend-like model and molecular gases by the Meek-like model, consistent with circumstantial evidence from previous studies.

\subsection{Townsend Breakdown and Paschen's Law}\label{townsendPaschen}

In a Townsend discharge, electrons moving from the cathode to anode acquire sufficient energy between collisions with gas atoms or molecules to cause secondary ionization from electron impact on neutral gas atoms or molecules.  The rate of such ionization processes per centimeter of transport is described by Townsend's first ionization coefficient, $\alpha(E)$, which is typically a strongly growing function of electric field $E$ under the field strengths relevant for breakdown.  The total number of electrons generated by a single electron emanating from the cathode is then $N_e=e^{\alpha d}$ where $d$  is the anode-to-cathode distance.

The production of secondary electrons also implies production of positive ions in equal numbers by charge conservation. These positive ions move toward the cathode, and when they arrive, may liberate additional electrons.  The number of liberated electrons per arriving positive ion was originally encoded in Townsend's second ionization coefficient, or the secondary emission coefficient, $\gamma_{se}$. The secondary emission coefficient is dependant upon the work function and therefore the material of the cathode, as well as the ionization potential of the surrounding gas \cite{alma992161253504911}.  It was subsequently realized that other secondary electron generating processes may also contribute to  $\gamma_{se}$~\cite{druyvesteyn1940mechanism} including photo-ionization or Penning ionization~\cite{csahin2010penning}, the latter being relevant in multi-component gas mixtures where excitation transfer from a high energy noble excimer to a molecular gas may be efficient.  The parameter $\gamma_{se}$ is thus considered an effective parameter determining how many secondary electrons are produced by the sum of these processes for each positive ion produced.

A breakdown will occur in a scenario where each cathode-originated electron produces a sufficient number of positive ions to generate at least one more electron from the cathode. Under such a condition, the current flow can sustain indefinitely. This criterion is encoded mathematically in the Townsend criterion, Eq.~\ref{townsendcriterion}:

\begin{equation}\label{townsendcriterion}
    \alpha(E) d=ln\left(1+\frac{1}{\gamma_{se}}\right).
\end{equation}

The first Townsend coefficient depends in a complex way on the micropysical behaviour of electrons traversing the gas.  To liberate a secondary electron, the electron must acquire sufficient energy between collisions to cause impact ionization.  The energy spectrum of swarm electrons depends on the cross sections for elastic and inelastic processes, and the cross section for impact ionization often has a complex energy dependence, especially for molecular gases.   Townsend introduced an empirical equation to describe the first ionization coefficient in several gases of experimental interest, Eq.~\ref{townsendapproximation} below \cite{meek1978electrical}. 

\begin{equation}\label{townsendapproximation} 
    \alpha(E) \sim Ape^{-\frac{Bp}{E}}.
\end{equation}

Combination of Eq.~\ref{townsendapproximation} with Eq.~\ref{townsendcriterion}, and making substitution $E=\frac{V_b}{d}$ leads to Paschen's Law, which models the breakdown voltage as a function of pressure $p$, gap distance $p$, empirical coefficients $A$, $B$ and $\gamma_{se}$:
\begin{equation}\label{paschen} 
    V_b=\frac{Bdp}{ln(Apd)-ln\left(ln\left(1+\frac{1}{\gamma_{se}}\right)\right)}.
\end{equation}

The A and B coefficients must be measured independently for each gas or gas mixture and many experimental values for different gases can be found in the literature, e.g Ref.~\cite{lieberman1994principles}. For attaching gases, another process is also active: negative ion formation resulting from capture of electrons on neutral gas molecules.  The Townsend criterion can be modified to include an additional coefficient, $\beta(E)$,  which represents the number of electrons lost through negative ion production per unit distance~\cite{uhm2000breakdown}. The resulting modified Townsend criterion and Paschen Law are shown below as equation \ref{attachmentcriterion} and \ref{attachmentpaschen} respectively \cite{uhm2000breakdown}. 

\begin{equation}\label{attachmentcriterion}
    (\alpha-\beta)d=ln\left(1+\frac{\xi}{\gamma_{se}}\right),\quad\quad\xi=1-\frac{\beta}{\alpha},
\end{equation}

\begin{equation}\label{attachmentpaschen}
    V_b=\frac{Bdp}{ln(Apd\xi)-ln\left(ln\left(1+\frac{\xi}{\gamma_{se}}\right)\right)}.
\end{equation}

For pure noble gases which have no attachment, $\xi=0$ and then equation \ref{attachmentpaschen} reduces to Eq. \ref{paschen}. It is notable that in these breakdown conditions, the breakdown voltage depends only on the combination $pd$ and not on $p$ and $d$ independently.

\subsection{Streamer Breakdown and the Meek-Raether Criterion}\label{meek}

A streamer breakdown is distinguished from a Townsend breakdown in that the electric field causing electron avalanche is dominated by the space charge associated with the avalanche itself, rather than the applied anode-to-cathode potential difference. The condition for formulation of a streamer is that the charge in an evolving avalanche must be sufficiently spatially concentrated that the $E$ field it generates is sufficient to sustain the avalanche.    

At each time during its evolution, the head of a streamer can be approximated as a sphere with radius $R(t)$.  This sphere of space charge generates an electric field $E'$, which can be obtained from Gauss's law. Given a known number of electrons $N_e=e^{\alpha(E) x}$, the field caused by space charge at the surface of the avalanche tip is:
\begin{equation}\label{spacechargefield}
   E' = \frac{\emph{e}N_e}{R^2} = \frac{\emph{e}\exp{\left(\alpha x\right)}}{R^2}.
\end{equation}
The electric field from space charge in the streamer exceeds the electric field applied between the electrodes when:
\begin{equation}\label{spacechargefield1}
   E' > E \quad \rightarrow \quad  \frac{\emph{e}\exp{\left(\alpha x\right)}}{R^2} > E.
\end{equation}
Raether~\cite{raether1939entwicklung} postulated that the value of $R$ for streamers could be associated with the diffusion radius of the advancing charge column, and subsequently measured $R$ in advancing streamers in air to be $R\sim0.013$~cm, consistent with this expectation.  Meek, and then Meek and Loeb~\cite{loeb1940mechanism,meek1940theory} used a similar principle to calculate that a self-sustained streamer breakdown would be expected when the value of $\alpha x$ satisfies Eq.~\ref{spacechargefield1} by the time $x=d$, which using the charge diffusion radius implies an order of magnitude prediction for the product $\alpha(E) d$ at the electric field strength where streamer formation begins:
\begin{equation}
[\alpha(E) d]_{crit}\sim 18-21.
\end{equation}
This is conventionally called the Meek-Raether criterion.  Although diffusion is central to Meek's mechanism, the diffusion constant does not feature in the numerical result, since it is eliminated in later stages of the derivation via use of the Einstein relation. Bazelyan and Raizer subsequently re-obtained the Meek-Raether criterion without invoking diffusion~\cite{bazelyan2017spark}, which shows that the Meek Raether criterion may be obtained from several different underlying mechanisms.  Subsequent work \cite{montijn2006diffusion}  has discussed corrections to the criterion from detailed modelling of charge transport around the streamer front.  

All derivations of this criterion involve a degree of ambiguity, encoded in the numerical value chosen for the critical value of $\alpha d$. However, the rapidly varying value of $\alpha(E)$ with electric field in the vicinity of a discharge over-voltage and the exponential dependence of the criterion on this number often means that the breakdown voltage prediction is not particularly sensitive to the critical value chosen, as we will soon verify.

\subsection{PyBoltz-Augmented Townsend and Meek Models}\label{pyboltz}

\begin{figure}[t]
\begin{centering}
\includegraphics[width=0.99\columnwidth]{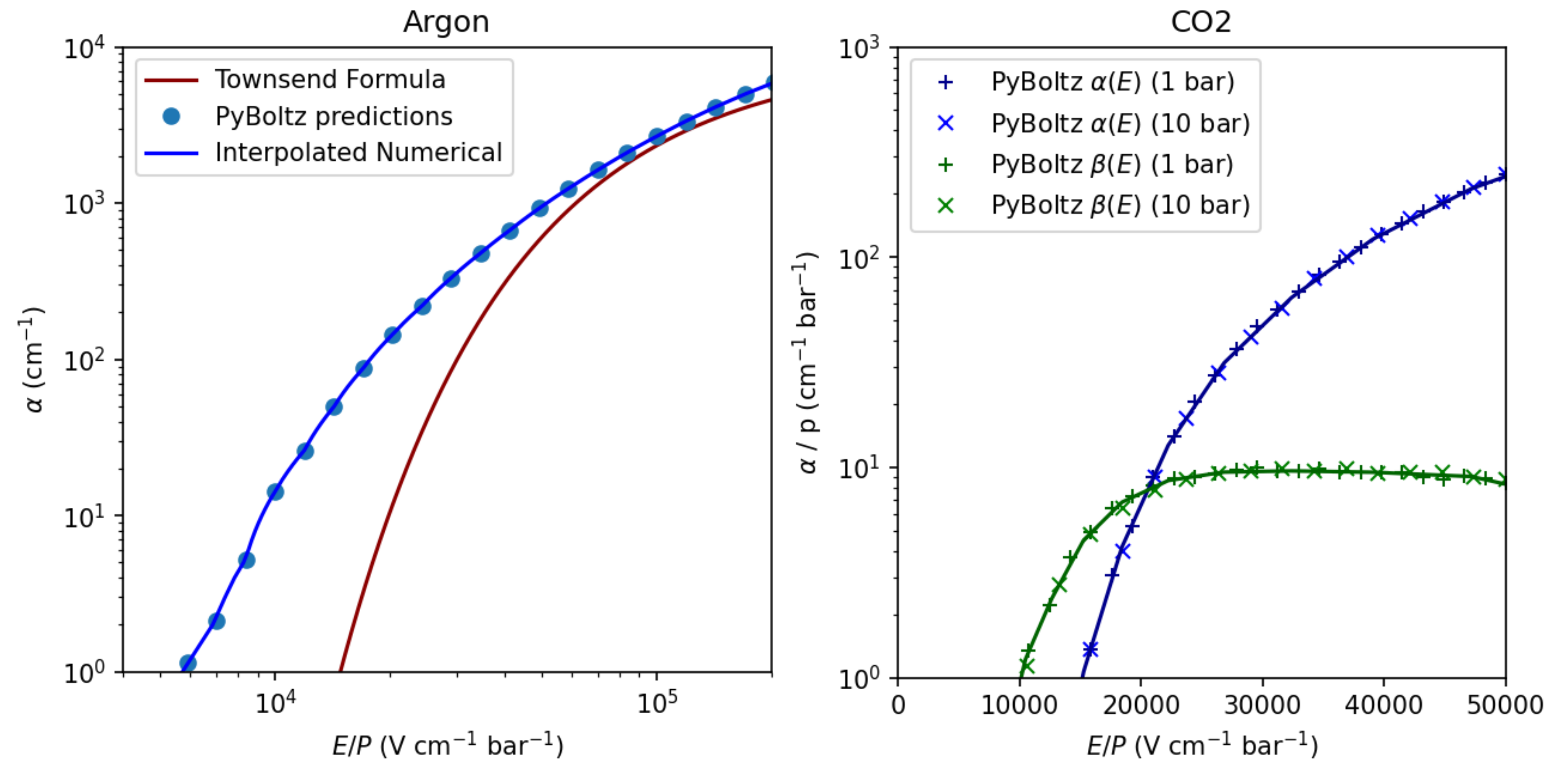}
\par\end{centering}
\caption{Left: comparison of the conventional Townsend approximation for $\alpha$ against the {\tt PyBoltz} microphysical prediction. The blue line shows the interpolated function which is inverted to use in solution of Eq.~\ref{eq:pyboltztownsend}.  Right: {\tt PyBoltz} calculations of reduced attachment and Townsend coefficients in CO$_2$ evaluated at 1 and 10 bar showing the expected pressure-dependent scaling. Also shown are the two interpolated functions used in numerical solution of Eq.~\ref{attachmentcriterion}.   \label{fig:PyBoltzTownsendExamples}}
\end{figure}

Our goal is to develop predictive models of high voltage breakdown in 1-10 bar pressures of gases of interest to time projection chamber detectors without arbitrary parameters for each gas. To this end, we have constructed modified Townsend and Meek models, which replace the empirical Townsend form for the $\alpha(E)$ Eq.~\ref{townsendapproximation} with one calculated from microscopic electron swarm simulations performed in the {\tt PyBoltz}~\cite{al2020electron} software package.  {\tt PyBoltz} performs Monte Carlo modelling of electron trajectories in gases and gas mixtures accounting for effects of elastic and inelastic scattering, ionization, attachment, predicting diffusion constants, drift velocities, and Townsend avalanche and attachment coefficients. Fig.~\ref{fig:PyBoltzTownsendExamples}, left shows as an example of the predicted functional dependence of  $\alpha(E)$ from {\tt PyBoltz} in comparison to the Townsend approximation with measured values of $A$ and $B$ from Ref.~\cite{lieberman1994principles} in pure argon. In this illustrative example we see clear differences between both the functional form and absolute value of the Townsend function between the conventional approximation and direct calculation.

The functional form of $\alpha(E)$ so obtained is then inserted into Eq.~\ref{townsendcriterion}, and inverted to find the breakdown voltage.  The resulting breakdown criterion is a modified Paschen-like law, with one free parameter $\gamma_{se}$, is:

\begin{equation}\label{eq:pyboltztownsend}
    V_b=d\alpha^{-1}\left[\frac{1}{d}ln\left(1+\frac{1}{\gamma_{se}}\right)\right].
\end{equation}

Since  $\gamma_{se}$ appears only inside a logarithm and $\alpha(E)$ varies rapidly with electric field in the vicinity of breakdown, even rather large variations $\gamma_{se}$ have a small impact on the predicted breakdown voltage.  A central value of $\gamma_{se}$=0.01 is assumed based on values that can be found in the literature, though a full order or magnitude around this number is considered when making our predictions,  0.0033<$\gamma_{se}$<0.033.  This has a small impact on the predicted breakdown voltages, only manifest at the lowest pressures.  One can also obtain an augmented Meek-like criterion by replacing $1+1/\gamma_{se}$ with a constant exp $\left([\alpha d]_{crit}\right)$. 

For gases with attachment, both $\alpha(E)$ and $\beta(E)$ are predicted, and Eq.~\ref{attachmentcriterion} is solved numerically.  Because in this case $\alpha$ and $\beta$ appear both inside  and outside the logarithm, there is not a simple expression like Eq~\ref{eq:pyboltztownsend} involving an inversion of $\alpha$ or $\beta$, though numerical solution for the value of $E$ that satisfies Eq.~\ref{attachmentcriterion} is straightforward. 

The mean energy of electron swarms in gases is predicted to depend on the ratio $E/P$ and not on $E$ or $P$ independently. This implies an expected scaling of various transport parameters with pressure. In particular, $\alpha$ and $\beta$ are both expected to scale as $\alpha/P$ and $\beta/P$ at fixed $E/P$. This allows for calculation of these two coefficients at one pressure in {\tt PyBoltz} followed by scaling to other operating pressures. The quality of the predicted scaling was checked with {\tt PyBoltz} for all the gases used in this study, with excellent agreement in all cases. Included in Fig.\ref{fig:PyBoltzTownsendExamples}, right is a comparison of  $\alpha/P$ and $\beta/P$ simulated at 1 and 10 bar demonstrating the accuracy of the predicted scaling behavior between pressures. All transport parameters used in this work are evaluated at 1 bar and scaled to the relevant working pressures.

It is notable that when $\alpha(E)$ scales with pressure in the expected way then $\alpha_p(E)=\alpha_0(E p_0/p) p/p_0$ and the breakdown voltage will be a universal function of $\chi=\frac{pd}{p_0}$, even without applying the Townsend approximation:
\begin{equation}\label{eq:pyboltztownsenduniv}
    V_b=\chi\alpha_0^{-1}\left[\frac{1}{\chi}ln\left(1+\frac{1}{\gamma_{se}}\right)\right].
\end{equation}
For attaching mixtures, in terms of the reduced Townsend and attachment coefficients the breakdown condition reads:
\begin{equation}\label{attachmentcriterionuniv}
    [\tilde{\alpha}_0-\tilde{\beta}_0]=\chi^{-1}\ln\left(1+\frac{1-\tilde{\beta}_0/\tilde{\alpha}_0}{\gamma_{se}}\right).
\end{equation}
Where the functions $\tilde{\alpha_0}$ and $\tilde{\beta_0}$ are the Townsend and attachment functions evaluated at $(V_b \chi^{-1})$:
\begin{equation}\label{tildefns}
    \tilde{\alpha_0}=\alpha_0[V_b \chi^{-1}],\quad \quad \tilde{\beta_0}=\beta_0[V_b \chi^{-1}].
\end{equation}
Solving Eq.~\ref{attachmentcriterionuniv} for V$_b$ numerically provides a value for the expected breakdown voltage, which will again scale as a universal function of $pd$.

\section{Experimental apparatus and method}\label{experimental apparatus}

\begin{figure}

    \centering
    \includegraphics[width = 0.25\linewidth]{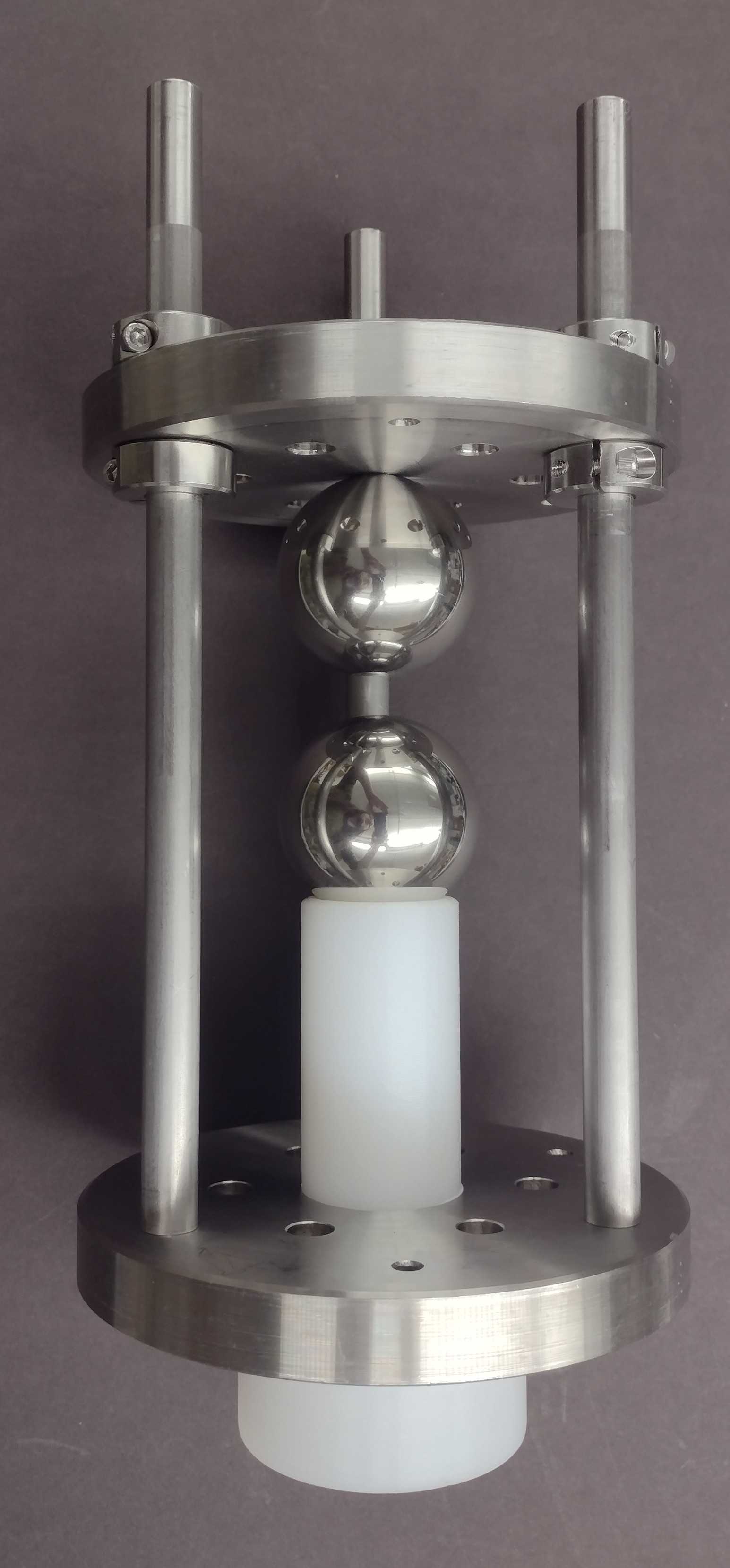}\includegraphics[width=0.7\linewidth]{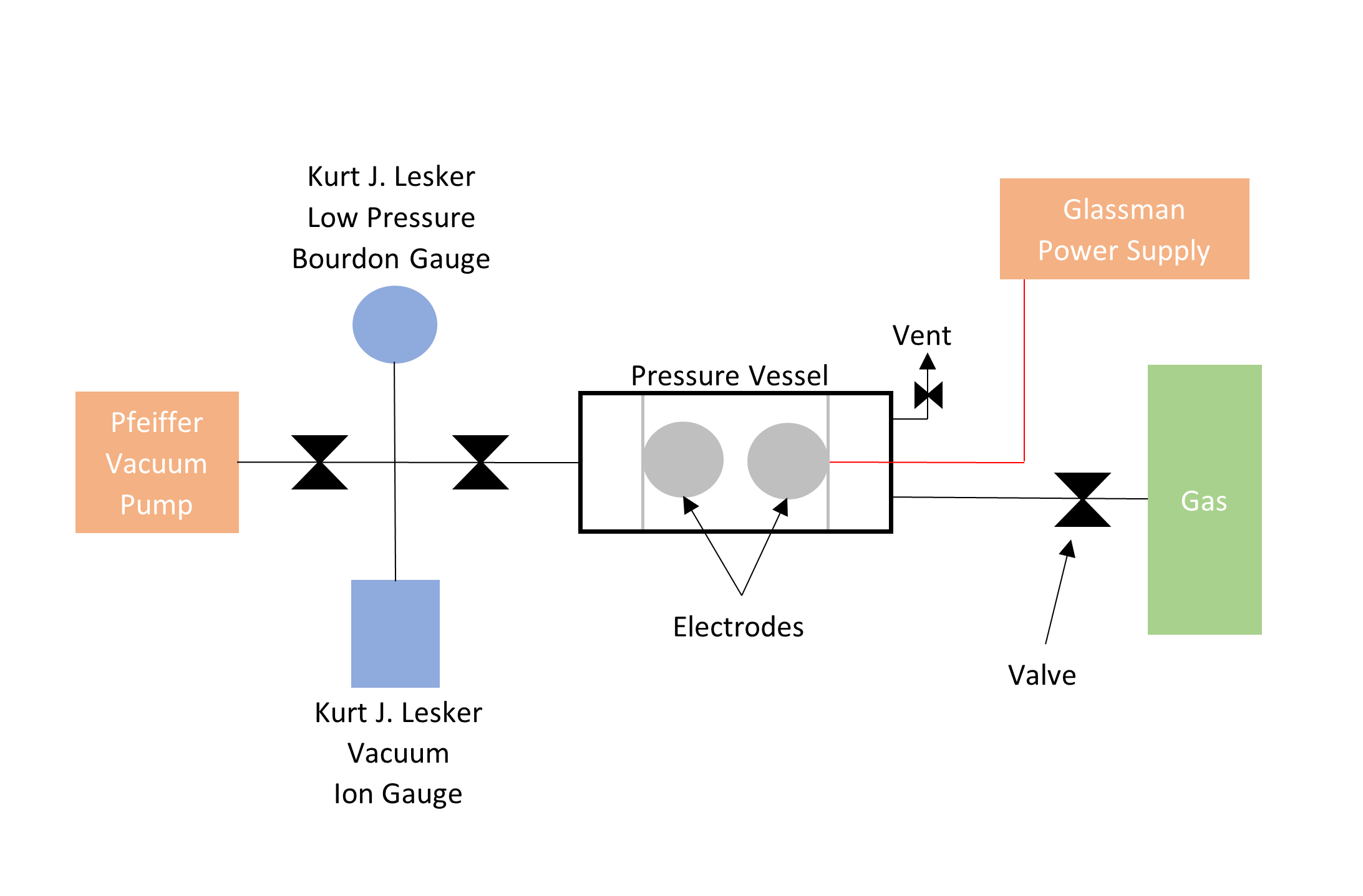}
\caption{Left: A photograph of the apparatus used in this study. Breakdowns are initiated between two polished spheres with an adjustable gap width, at voltages of up to 50 kV. Right: Block diagram of the experiment showing key system components.}
\label{fig:Setup}
\end{figure}

To perform the breakdown measurements we designed and assembled a test stand with accurately positionable spherical electrodes operable at pressures between 0.1 and 10 bar.  Fig.~\ref{fig:Setup}, left, shows the device, consisting of two 50~mm diameter spherical 316 stainless steel electrodes that are attached to two large stainless steel disks.  The disks fit snuggly into a pressure vessel and locate the electrodes at its center, maintaining at least 4~cm each side between the spheres and the 6~inch diameter inner wall.  Electrode distances between 0.1 and 10~mm were used. The upper gap limit was chosen to ensure the gap size was always small compared to the sphere radii and sphere-to-wall or rod distances, which at larger scales could introduce unacceptable field distortions.

The cathode sphere connects to a custom-manufactured high voltage feed-through that sits on a center flange and connects through a metal contact running through the HDPE insulator seen at the bottom of Fig.~\ref{fig:Setup}, left. The feed-through was made by cryo-fit of polyethylene into steel tube and performs to our specification of holding at least 55~kV high voltage, 10 bar operating pressure and evaccuability to better than 10$^{-6}$ Torr.  The high voltage cathode disk is stationary while the ground disk slides along the guide rods to set the gap spacing. The position is set by placing precision thickness gauges within the gap between the spheres and then locking the shaft collars on each side of the disk preventing movement.  The disks are not fixed to the pressure vessel and can slide over it, so the breakdown gap distance is not affected by any elastic deformations when the system is pressurized.

The high voltage power supply used was a Glassman KT-100 series power supply capable of delivering a maximum voltage of 100 kV and a maximum current of 20 milliamps. The  power supply has a precision of $\pm$ 50 Volts from the readout panel, however an improved precision of  $\pm$ 5 Volts can be obtained via multimeter on the monitoring port.  We set a maximum operating voltage of 55~kV for this study, in order to protect the feed-through.  The experiment was performed over  two distinct ranges of pressure, 1-10 bar and 0.1-1 bar, and to measure pressure we used two separate Bourdon pressure gauges, one for each range. The lower range gauge had a range of 0.1-1 bar of absolute pressure and a precision of $\pm$ 0.005 bar. The upper range gauge had a range of 0-20 bar of gauge pressure and a precision of $\pm$ 0.25 bar.

Before each run the vessel was evacuated overnight, and the experiments were started only when residual gas pressures of order $10^{-6}$ mbar were obtained, as read from an ion gauge on the vacuum line. The vessel was isolated from vacuum using a Carten valve and filled with test gas to the specified pressure. Every gas was of a minimum 99.999\% purity, including the individual gases used in the gas mixtures which had an uncertainty of $\pm$ 0.01\% on the mix ratio. Gases and mixtures that were compatible were filled through a SAES cold getter for initial purification. We made a comparison run in pure argon from the gas bottle without purification and found equivalent breakdown voltages, so even in the case of molecular gas mixtures which could not be purified, the raw gas purity is expected to be sufficiently high not to compromise the dielectric strength measurements.

Once the vessel was filled to the desired pressure, the voltage was steadily increased until a sharp increase in current occurred, tripping the over current protection on the power supply, which was taken as evidence of a high voltage breakdown across the gap. This was repeated for at least three breakdown measurements at each pressure increment and a range of pressures for each gas and electrode separation distance.  To test for possible hysteresis, comparative measurements were made with increasing and then decreasing pressure increments for argon gas, with very similar results in both cases. The difference of average magnitude 6\% is added in quadrature as a small uncertainty contribution in the final error bars.

\begin{figure}[t]
\includegraphics[width = \linewidth]{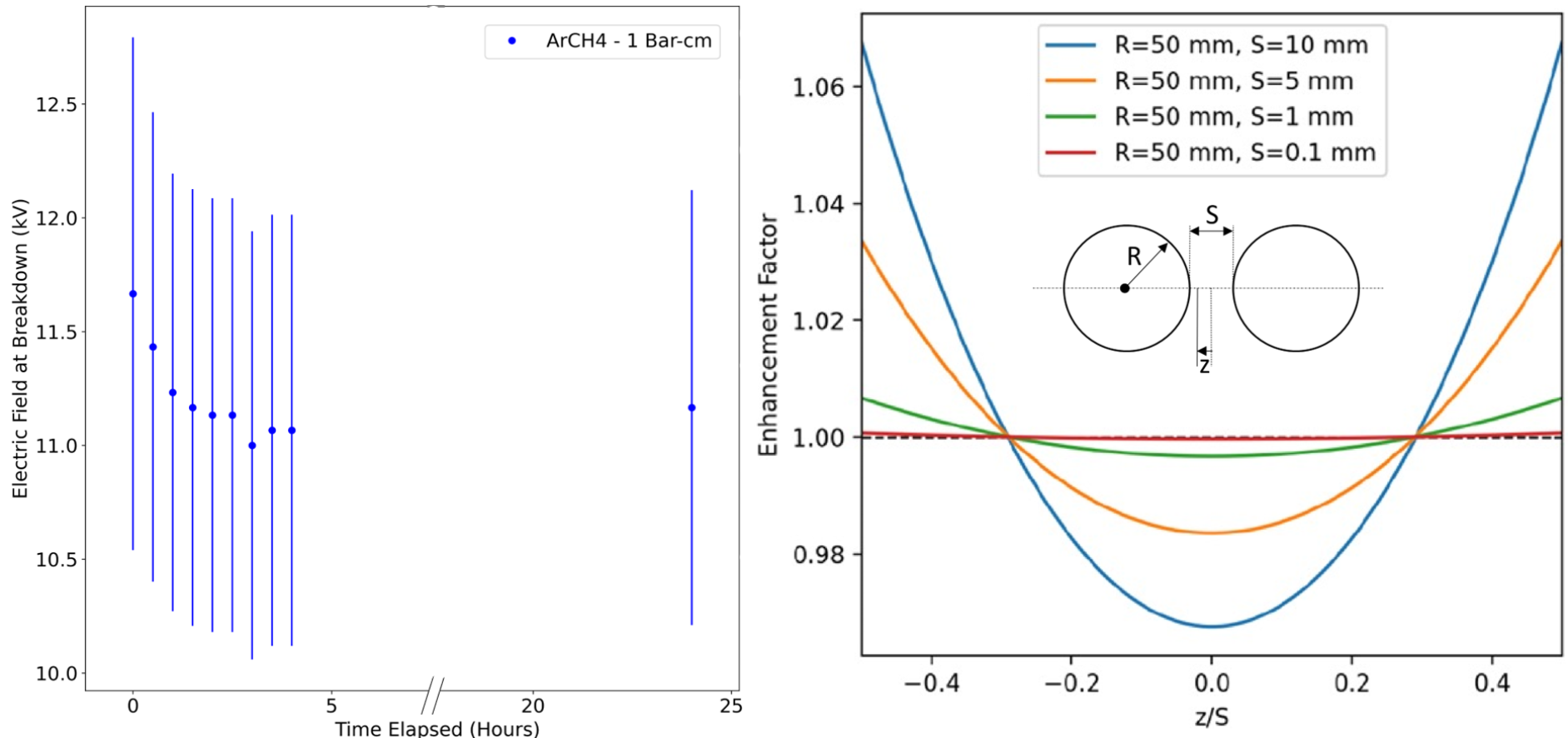}
\caption{Left:Breakdown Voltage as a function of the time elapsed after pressurizing to 1 bar of Ar-CH$_4$; Right: calculation of enhancement factors of electric field relative to parallel plate geometry as a function of position in the gap between the two spheres $z$ for fixed values of radius $R$ and gap size $S$ used in this study}
\label{fig:systplots}
\end{figure}

Since our device does not incorporate continual circulation and re-purification, to test for time dependence of the breakdown strength due to out-gassing  of water during the run a time dependent study was undertaken. A set of breakdown measurements were taken at one bar pressure immediately after filling the vessel and again at time intervals up to 24 hours. The results of this study can be seen in Figure \ref{fig:systplots} with data taken at every 30 minute interval and the last data point taken at 24 hours.  An $\cal{O}$(5\%) deviation is observed, taking place over around five hours.  Since one run typically takes 1-2 hours starting from clean gas in an evacuated vessel, this degree of variation is small. It is and also somewhat degenerate with the previously described hysteresis uncertainty, though is conservatively included in the final error bars as an uncorrelated contribution.

A systematic uncertainty is applied to account for the non-planar nature of the electrodes leading to  breakdown occurring in a region of slowly varying electric field rather than a uniform field configuration, as in the idealized Townsend discharge.  We performed a calculation to determine the ratio of field strength of spherical electrodes to planar electrodes (which we call the 'enhancement factor') following the methods of Ref~\cite{chaumet1998electric}. The determined enhancement factor is shown in Fig~\ref{fig:systplots}, right.  The largest value of the enhancement factor in the gap is taken as a systematic uncertainty on the applied voltage leading to breakdown for each geometry.  In practice it is expected that the breakdown is initiated in the highest field region and so we expect the true values to be near the upper part of this allowed range, which has a span of around $\pm$ 6\%.

Each data point in our presented results represents the mean of at least three breakdown measurements at each pressure point, with the standard deviation added in quadrature to the systematic uncertainties described above to form the error bar.  In all cases the statistical spread is dominant over all of the systematic uncertainties.

\section{Results and Discussion}\label{results}

\begin{figure}[t]
\includegraphics[width = \linewidth]{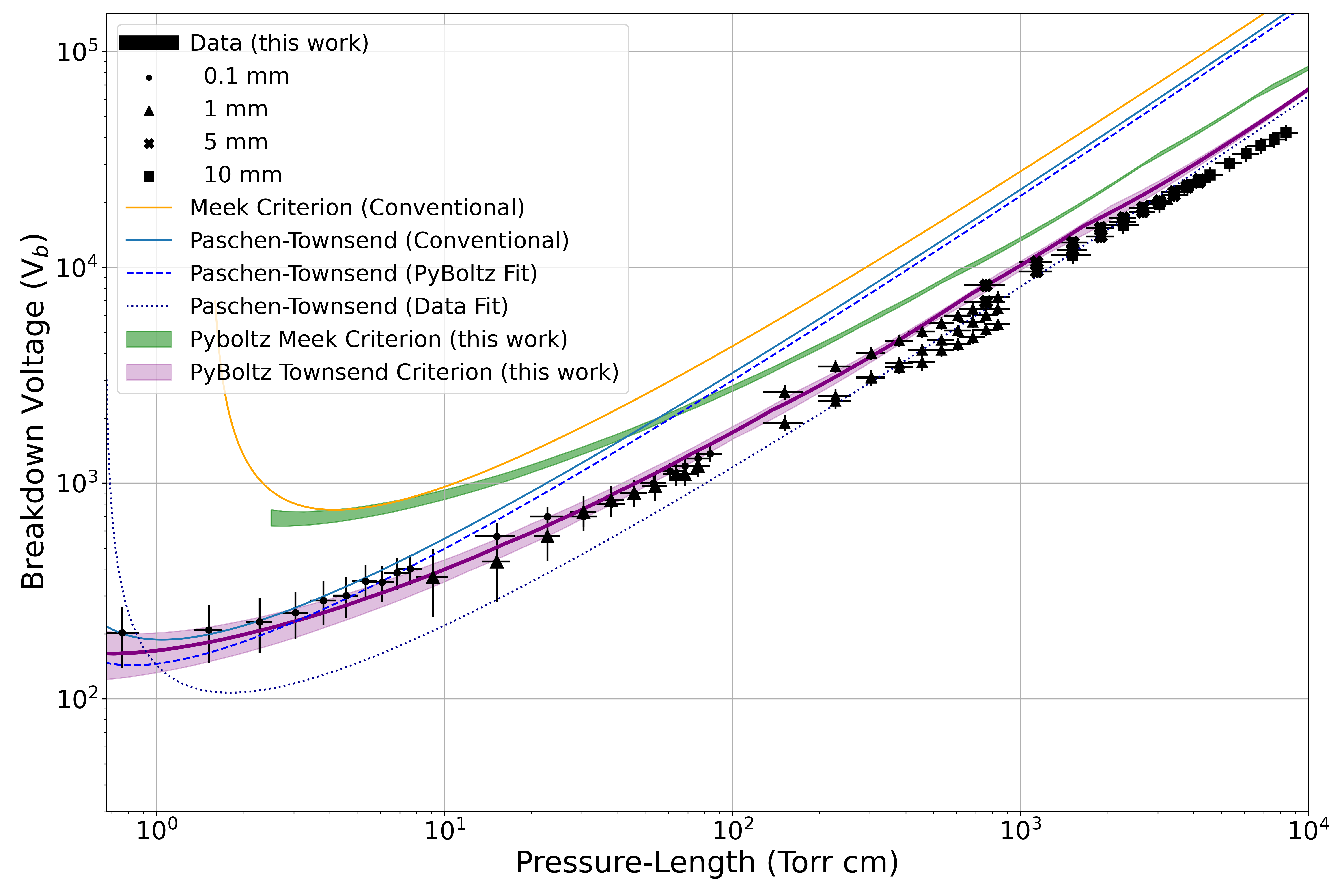}
\caption{Argon breakdown voltage vs. Pressure-Length data plotted alongside the theoretical curves}
\label{argonplot}
\end{figure}

Data was taken at varying gap distances to span the full range of $pd$ values of interest for several gas mixtures of interest to future neutrino detectors: pure argon, pure xenon, pure CO$_2$, pure CF$_4$, Ar-CO$_2$ (90\%/10\%),  Ar-CH$_4$ (90\%/10\%) and Ar-CF$_4$ (99\%/1\%).  The largest dataset  was accumulated with pure argon gas, with gap spacings 0.1~mm to 10~mm in order to make detailed comparisons between available high voltage breakdown models.  The resulting dataset can be seen in Fig.~\ref{argonplot} alongside various theoretical curves.  We show the traditional Paschen curve with the Townsend approximation for $\alpha(E)$ with $A$ and $B$ taken from Ref~\cite{alma992161253504911}, which fits the data well at lower pressures but diverges rapidly at higher ones, as well as the Meek criterion with conventional values for $A$ and $B$ which does not fit the data well anywhere. 

We also show two intermediate Paschen-Townsend curves, which have the traditional form but re-fitted parameters. The first has $A$ and $B$ in the Townsend approximation for $\alpha(E)$, Eq.~\ref{townsendapproximation},  fitted to {\tt PyBoltz} predictions over the $pd$ range of interest.  This prediction represents, in a sense, the best theoretically informed version of the traditional Paschen curve. This theoretical description shows similar deficiencies to the original Paschen-Townsend curve, agreeing well at low pressure and poorly at high pressure. It thus appears that the primary deficiency of the Paschen-Townsend model lies in the  form of the empirical parameterization Eq.~\ref{townsendapproximation} and not only its parameters, since even after making our best theoretical prediction of the free parameters it does not match experiment well over all $pd$ values.  We also show a Paschen-Townsend curve with $A$ and $B$ fitted to our experimental data directly.  Naturally, this model fits more closely to the data used to constrain it, though the values of $A$ and $B$ that emerge from the fit are far from the values typically measured at low $pd$, and their interpretation in this context is questionable.  In this fit $\gamma_{se}$ was fixed to 0.01 with $A$ and $B$ free.  Allowing $\gamma_{se}$ as a further free parameter and fitting a three-parameter Townsend-Paschen model to our data yields $\gamma_{se}=0.02$ and an almost indistinguishable Paschen-Townsend curve (so much so that we omit it from Fig~\ref{argonplot} to avoid over-cluttering). Values of the fitted $A$ and $B$ for each of these models are tabulated in Table~\Ref{tabvalues}.

Also shown on Fig.~\ref{argonplot} are the two models that dispense with $A$ and $B$ and represent central theoretical products of this work: the {\tt PyBoltz}-augmented Townsend and Meek models for breakdown voltage, calculated per the methods of Sec.~\ref{pyboltz}.  The shaded region around the Townsend-like model covers an order of magnitude of realistic values of $\gamma_{se}$ as described in Sec.~\ref{pyboltz}. The shaded region around the Meek-like model covers the range of viable $[\alpha d]_{crit}$ values 18 to 21.  At the lowest $pd$, the region of smaller gap sizes and lower pressures where the Paschen curve is most frequently studied, this prediction is in good agreement with the conventional Paschen-Townsend description.  The model diverges from the other descriptions at higher pressures but maintains a good match to data, out-performing all of the previous models in terms of its ability to describe the data over the full range of $pd$ values of this study. The weak (logarithmic) dependence of the breakdown voltage upon the only unknown parameter $\gamma_{se}$ implies a high degree of predictivity, without sensitive dependence of breakdown voltage on {\em{a-priori}} unknown empirical parameters. The {\tt PyBoltz}-augmented Meek model is not a good fit at lower $pd$ values but tends toward the {\tt PyBoltz}-augmented Townsend prediction at higher pressures.

\begin{table}[t]
\begin{centering}

\begin{tabular}{llll}
\hline 
Source & A & B & $\gamma_{se}$\tabularnewline
\hline 
\hline 
Conventional value \cite{alma992161253504911}& (12) & (180) & (0.01) \tabularnewline
\hline 
Fit to {\tt PyBoltz} $\alpha(E)$ & 15 & 173 & (0.01) \tabularnewline
\hline 
Fit to data $V_{B}(pd)$:  AB($\gamma_{se}$) & 3.6 & 52 & (0.01) \tabularnewline
\hline 
Fit to data $V_{B}(pd)$:  AB$\gamma_{se}$ & 3.0 & 52 & 0.02 \tabularnewline
\hline 
\end{tabular}
\par\end{centering}
\caption{$A$, $B$ and $\gamma$ coefficients from various descriptions of the breakdown voltage in terms of the Paschen-Townsend parameterization. Numbers in (brackets) are fixed constants, whereas those without are fit results. The first row shows the conventional values from Ref~\cite{alma992161253504911}.  The second row shows the result of fitting the Townsend approximation \ref{townsendapproximation} to {\tt PyBoltz} predictions of $\alpha(E)$ for argon. The third and forth rows show the result of fitting the Paschen curve to data directly, either with or without $\gamma_{se}$ as a free parameter; in practice these two models are nearly indistinguishable.  None of these models fits the data well over all $pd$ values as well as the {\tt PyBoltz}-Townsend model, which predicts $\alpha(E)$ directly rather than relying on $A$ and $B$ coefficients. \label{tabvalues}  }
\label{AAndBTable}
\end{table} 

\begin{figure}[t]
\includegraphics[width = \linewidth]{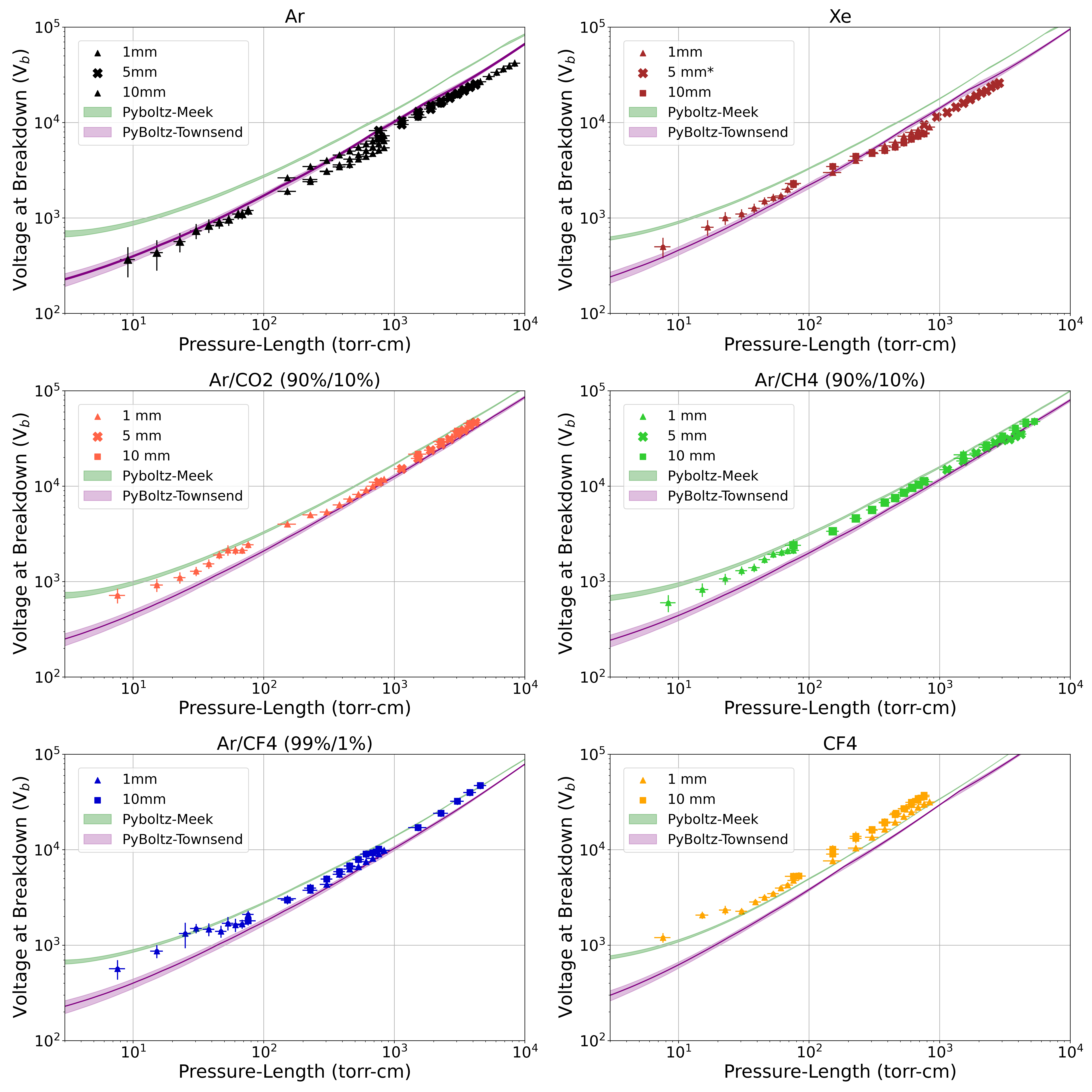}
\caption{Breakdown Data for Various Gases and Gas Mixtures plotted alongside the {\tt PyBoltz} Predictions for each gas. *5mm Xe data presented from a functionally equivalent device in Ref.~\cite{rogers2018high}}
\label{gasplots}
\end{figure}

\begin{figure}[t]
\begin{centering}
\includegraphics[width =\linewidth]{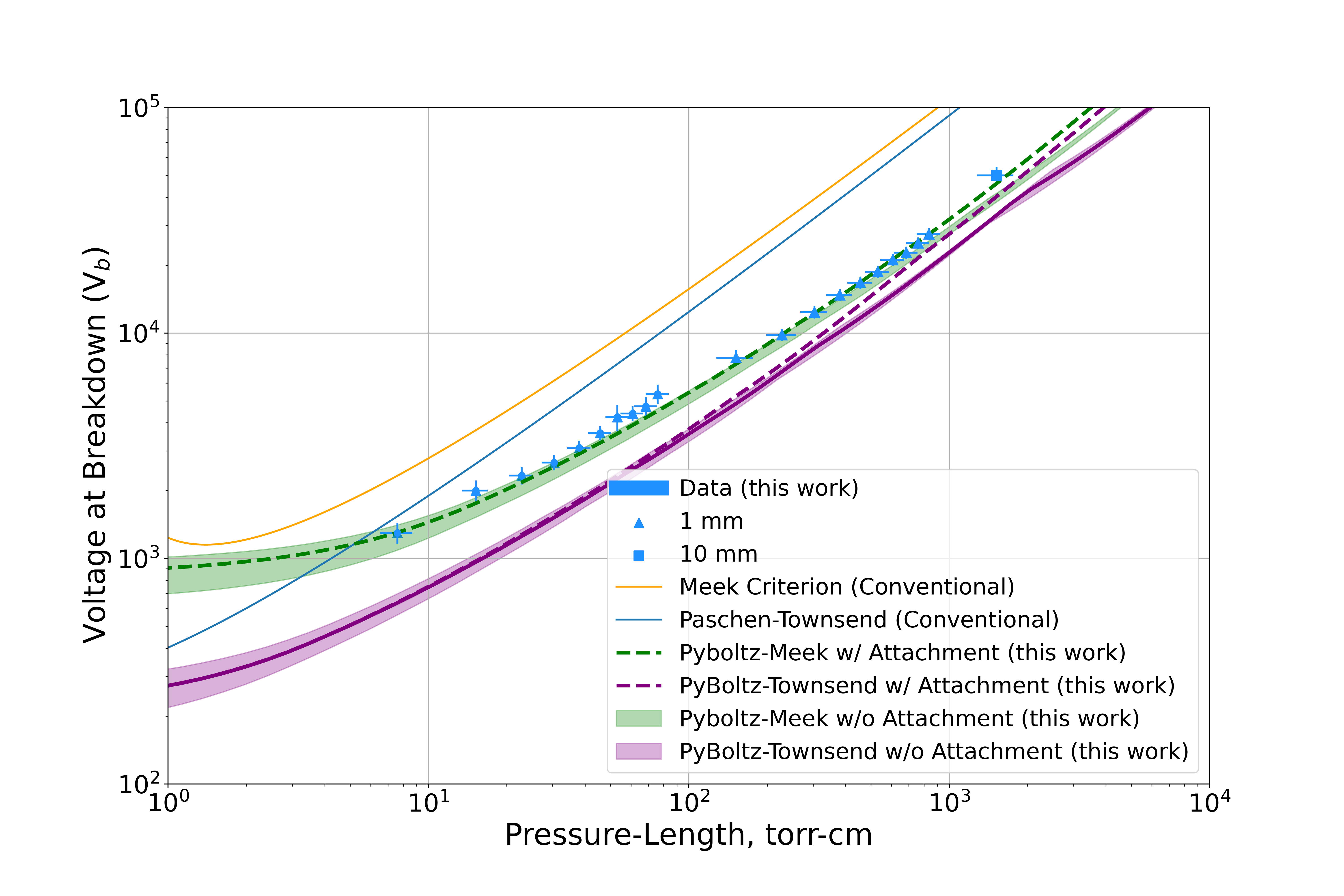}
\par\end{centering}
\caption{CO$_2$ breakdown voltage vs. Pressure-Length data plotted alongside the theoretical curves}
\label{fig:CO2plot}
\end{figure}

Fig~\ref{gasplots} shows similar comparison for argon plus five other gases and gas mixtures studied in this work, compared to the theoretical {\tt PyBoltz}-augmented Townsend and Meek predictions.  The {\tt PyBoltz} calculations are rather predictive in all cases, especially at higher pressures with small offsets between the curves and data for every gas except Argon.   In xenon gas, once again the {\tt PyBoltz}-augmented Townsend model appears very accurate at all values of $pd$.   The pure molecular gases CF$_4$ and CO$_2$ (shown separately in Fig~\ref{fig:CO2plot}) appear to fit significantly better to {\tt PyBoltz}-augmented Meek predictions rather than {\tt PyBoltz} augmented Townsend ones, however.  The noble-molecular gas mixtures appear to be intermediate, relatively well described by either model, though with some tendency towards the Meek-Raether model, as may be expected in quenched gas mixtures.

Since the {\tt PyBoltz}-augmented Townsend model clearly under-predicts the breakdown strength at the lowest $pd$ values for photon-quenched gases despite having being validated very well over all $pd$ values in nobles, we may postulate that the dielectric breakdowns in the noble gases are more Townsend-like in nature whereas the dielectric breakdowns in the molecular gases are more streamer-like in this pressure range. Before discussing this hypothesis further, we consider two other possibilities.  First, we note that the largest deviations from Townsend-like behaviour are observed in CO$_2$ and CF$_4$, which both have relatively strong attachment, unlike argon and xenon.  We should consider whether the method of incorporating attachment into the Townsend formalism is sufficient, since it is based on steady-state discharge arguments, and the sparks in this system are clearly transient events.  Comparison of the with- and without-attachment models for CO$_2$ shown in Fig.~\ref{fig:CO2plot} demonstrates that the effects of attachment in the breakdown criterion primarily emerge only at higher $pd$ values, and thus are not likely to explain the Townsend model deficiencies at lower $pd$.  A further consideration is that the molecular gases are clearly more complex to model in swarm simulations, in particular their scattering and ionization cross sections may be less well characterized than for pure argon or xenon. On the other hand, both CO$_2$ and CF$_4$ are relatively widely used gases in particle detectors~\cite{csahin2016systematic}, and both sets of cross sections used in this work have a five-star demarcation in the {\tt MagBoltz}~\cite{biagi1999monte} package, the highest quality rating.  It thus appears unlikely that they are so deficient as to cause the observed effects. 

We thus consider the plausible yet speculative explanation, that the Townsend-to-streamer transition occurs at lower pressure-distance values for CO$_2$ and CF$_4$ and Ar-based mixtures than for pure Ar and Xe gases, and indeed the discharges in noble gases should be considered as Townsend-like whereas those in insulating or quenched gases should be considered as Meek-like.  In dry air, the transition between Townsend and streamer-like discharge has been reported to occur at a $pd$ value above around 5000 Torr~cm~\cite{davies1965electrical,allen1964correlation}, the upper end of the values explored in this work.  It is therefore very reasonable to consider transitions between these two descriptions would occur in this regime for the various gases tested.  That the quenched gases may develop streamers at lower pressures than the noble and predominantly noble gases also appears consistent with previous results.  For example, Ref~\cite{hodges1985probability} studies time evolution of discharges at pressures of up to 1~bar in various gases.  There, argon gas showed Townsend-like behaviour of the time profile for all tested values of $pd$, whereas a transition from Townsend-like to streamer-like behaviour was observed in the molecular gases SF$_6$ and CCl$_2$F$_2$ at pressures below 1~bar.  This appears to be in good circumstantial agreement with our observations of a Meek-like breakdown criterion in photon quenched and insulating gases and a more Townsend-like one in noble gases.  

While certainly of fundamental interest, it is notable that this distinction at low $pd$ is of limited pragmatic interest for time projection chambers applications. The proposed uses, especially of pure CO$_2$ and pure CF$_4$ as gas-phase insulators, involve standing off high voltages at large gap distances and high pressures, for which the largest $pd$ values probed in this work are most relevant. In this regime the Townsend and Meek predictions for breakdown voltage become equivalent, and our model is predictive independently of the critical value $[\alpha d]_{crit}$ or $\gamma_{se}$.

\section{Conclusions\label{sec:Conclusions}}

We have studied a suite of gases being actively considered as working media for near-future high pressure gas time projection chamber detectors in neutrino physics.  Due to its low breakdown voltage and lack of photon quenching, pure argon gas is not a common working medium for large gas-phase time projection chambers, though was studied as a benchmark for comparison against historical data and models. Doped argon gas mixtures are of great interest, and in particular all of 90-10 argon-CH$_4$, 90-10 argon-CO$_2$ and 99-1 argon-CF$_4$ are being actively considered as working media for a high pressure gas near detector for the DUNE experiment.  Surrounding this active volume may be a volume of insulating gas and in each case the minority component of the drift gas would be a favorable choice, to minimize the impact of possible (though undesirable) diffusive leaks between inner volumes.  To this end we have also studied pure CO$_2$ and CF$_4$. Pure CH$_4$ is not an appealing insulating gas due to its flammability, and has not been studied. Finally we also studied pure xenon, the active medium of high pressure gas time projection chambers searching for neutrinoless double beta decay in $^{136}$Xe.

In all cases, we observe deviations from Paschen-like pressure scaling at high pressures, which can be attributed to imperfection of the empirical Townsend approximation for describing the dependence of the Townsend coefficient $\alpha(E)$ upon electric field.  A new model, which uses the {\tt PyBoltz} electron swarm simulation code to predict the Townsend $\alpha$ and $\beta$ coefficients was used to derive an augmented Townsend or Meek discharge model that is found to be highly predictive for the majority of the gases considered in this study.  Both {\tt PyBoltz}-augmented Townsend-like and Meek-like models are strongly predictive for all gases at large $pd$ values, unlike any of the conventional models.  

\begin{table}[b]
\begin{centering}

\begin{tabular}{llllllll}
\hline
\multicolumn{8}{c}{Projected Breakdown Voltage at 10~bar, 1~cm (kV)} 
\tabularnewline
\hline
\hline 
 & Ar & Xe & Ar-CF$_4$ & Ar-CH$_4$ & Ar-CO$_2$ & CO$_2$ & CF$_4$ \tabularnewline
\hline 
Townsend &{\bf 52.6} & {\bf 75.4} & 61.7 & 63.9 & 68.6 & 129.5 & 179.7 \tabularnewline
\hline 
Meek & 69.9 & 98.9 & 72.1 & 80.3  & 87.3 & {\bf 171.2} & {\bf 212.2}\tabularnewline
\hline
\end{tabular}
\par\end{centering}
\caption{Projected breakdown voltage of several gases at a benchmark point of 10~bar pressure, 1~cm gap distance. Based on the observed trends the bold values are the recommended projections for the pure gases.}
\label{ProjectedBreakdown}
\end{table}

For noble gases, the {\tt PyBoltz}-augmented Townsend model is predictive for all $pd$ values investigated, whereas for the pure molecular gases CF$_4$ and CO$_2$ the {\tt PyBoltz}-augmented Townsend model under-predicts the breakdown voltage data by factors of up to two.  A {\tt PyBoltz}-augmented Meek criterion provides a good description of the data for these gases.   We consider this as circumstantial evidence that a transition from Townsend-like to Meek-like discharges is present at lower pressures than in the noble gases, and this explanation also appears consistent with other reported studies.  While plausible, we do not overlook that since microphysics of CO$_2$ and CF$_4$ are rather more complex than the pure noble gases, the observed differences may also be associated with imperfect collision or impact ionization cross sections in the {\tt PyBoltz} code. At the higher pressures and larger gap distances which are of the most pragmatic interest for this work, either  the Townsend-like or Meek-like models become sufficiently accurate for projecting breakdown voltages in all cases, including for molecular gases CO$_2$ and CF$_4$ and gas mixtures.

Due to limitations imposed by our high voltage feed-through, we were not able to induce breakdown in all gases at 10~bar pressure and 1~cm distance below the maximum safe operating voltage of 55~kV.  In all such cases (all gases except argon) we conclude a breakdown field of at least 55~kV over 1~cm is sustainable at 10 bar.  Since the data and models agree very well at the higher pressures and gap sizes, we can also extrapolate the breakdown voltages observed at the highest pressures and gap distances to the benchmark point of 1~cm and 10~bar. Table~\ref{ProjectedBreakdown} presents these projections for both models.   Argon-based gas mixtures are projected to support electric fields of $\sim$90~kV/cm over 1~cm-scale gaps; the insulating gases CO$_2$ and CF$_4$ are projected to support $\sim$170~kV/cm and $\sim$210~kV/cm respectively.  These dielectric strengths are very promising for the proposed applications as active gases and insulating gases in large, high pressure gas time projection chambers.

\section*{Acknowledgements}
We thank Jen Raaf, Alysia Marino, Alan Bross, Diego Gonzalez Diaz for their comments and encouragement throughout this work.  The UTA group acknowledges support from the University of Texas at Arlington and Department of Energy under Early Career Award number DE-SC0019054, and by Department of Energy Award DE-SC0019223.

\printbibliography

@book{alma992161253504911,
    author = {Howatson, A. M.},
    address = {Oxford},
    edition = {[1st ed.].},
    language = {eng},
    lccn = {63022500 //r852},
    publisher = {Pergamon Press},
    series = {The Commonwealth and international library. Applied electricity and electronics division ; v. 8},
    title = {An introduction to gas discharges},
}

@book{meek1978electrical,
  title={Electrical breakdown of gases},
  author={Meek, John M and Craggs, John Drummond},
  year={1978}
}

@article{montijn2006diffusion,
  title={Diffusion correction to the Raether--Meek criterion for the avalanche-to-streamer transition},
  author={Montijn, Carolynne and Ebert, Ute},
  journal={Journal of Physics D: Applied Physics},
  volume={39},
  number={14},
  pages={2979},
  year={2006},
  publisher={IOP Publishing}
}

@article{csahin2010penning,
  title={Penning transfer in argon-based gas mixtures},
  author={{\c{S}}ahin, {\"O} and Tapan, I and {\"O}zmutlu, Emin N and Veenhof, R},
  journal={Journal of Instrumentation},
  volume={5},
  number={05},
  pages={P05002},
  year={2010},
  publisher={IOP Publishing}
}

@article{druyvesteyn1940mechanism,
  title={The mechanism of electrical discharges in gases of low pressure},
  author={Druyvesteyn, MJ and Penning, Fi M},
  journal={Reviews of Modern Physics},
  volume={12},
  number={2},
  pages={87},
  year={1940},
  publisher={APS}
}

@article{biagi1999monte,
  title={Monte Carlo simulation of electron drift and diffusion in counting gases under the influence of electric and magnetic fields},
  author={Biagi, SF},
  journal={Nuclear Instruments and Methods in Physics Research Section A: Accelerators, Spectrometers, Detectors and Associated Equipment},
  volume={421},
  number={1-2},
  pages={234--240},
  year={1999},
  publisher={Elsevier}
}

@article{allen1964correlation,
  title={Correlation of the formative time lags with the light emitted from spark discharges},
  author={Allen, KR and Phillips, K},
  journal={Proceedings of the Royal Society of London. Series A. Mathematical and Physical Sciences},
  volume={278},
  number={1373},
  pages={188--213},
  year={1964},
  publisher={The Royal Society London}
}

@article{hodges1985probability,
  title={Probability of electrical breakdown: Evidence for a transition between the Townsend and streamer breakdown mechanisms},
  author={Hodges, RV and Varney, RN and Riley, JF},
  journal={Physical Review A},
  volume={31},
  number={4},
  pages={2610},
  year={1985},
  publisher={APS}
}

@article{davies1965electrical,
  title={Electrical breakdown of air at high voltages},
  author={Davies, WEVJ and Dutton, J and Harris, FM and Jones, F Llewellyn},
  journal={Nature},
  volume={205},
  number={4976},
  pages={1092--1093},
  year={1965},
  publisher={Nature Publishing Group}
}

@article{chaumet1998electric,
  title={Electric potential and field between two different spheres},
  author={Chaumet, PC and Dufour, JP},
  journal={Journal of electrostatics},
  volume={43},
  number={2},
  pages={145--159},
  year={1998},
  publisher={Elsevier}
}

@article{loeb1940mechanism,
  title={The mechanism of spark discharge in air at atmospheric pressure. I},
  author={Loeb, Leonard B and Meek, John M},
  journal={Journal of applied physics},
  volume={11},
  number={6},
  pages={438--447},
  year={1940},
  publisher={American Institute of Physics}
}

@book{bazelyan2017spark,
  title={Spark discharge},
  author={Bazelyan, Eduard Meerovi{\v{c}} and Raizer, Yu P},
  year={2017},
  publisher={Routledge}
}

@article{meek1940theory,
  title={A theory of spark discharge},
  author={Meek, JM},
  journal={Physical review},
  volume={57},
  number={8},
  pages={722},
  year={1940},
  publisher={APS}
}

@article{raether1939entwicklung,
  title={Die entwicklung der elektronenlawine in den funkenkanal},
  author={Raether, H},
  journal={Zeitschrift f{\"u}r Physik},
  volume={112},
  number={7},
  pages={464--489},
  year={1939},
  publisher={Springer}
}

@article{lieberman1994principles,
  title={Principles of plasma discharges and materials processing},
  author={Lieberman, Michael A and Lichtenberg, Allan J},
  journal={MRS Bulletin},
  volume={30},
  number={12},
  pages={899--901},
  year={1994}
}

@article{Nygren:1974nfi,
    author = "Nygren, D. R.",
    editor = "Kadyk, J and Nygren, D. and Wenzel, W. and Winkelmann, F.",
    title = "{The Time Projection Chamber: A New 4 pi Detector for Charged Particles}",
    reportNumber = "PEP-0144",
    journal = "eConf",
    volume = "C740805",
    pages = "58",
    year = "1974"
}

@article{Gonzalez-Diaz:2017gxo,
    author = "Gonzalez-Diaz, D. and Monrabal, F. and Murphy, S.",
    title = "{Gaseous and dual-phase time projection chambers for imaging rare processes}",
    eprint = "1710.01018",
    archivePrefix = "arXiv",
    primaryClass = "physics.ins-det",
    doi = "10.1016/j.nima.2017.09.024",
    journal = "Nucl. Instrum. Meth. A",
    volume = "878",
    pages = "200--255",
    year = "2018"
}

@article{NEXT:2019gtz,
    author = "Ferrario, P. and others",
    collaboration = "NEXT",
    title = "{Demonstration of the event identification capabilities of the NEXT-White detector}",
    eprint = "1905.13141",
    archivePrefix = "arXiv",
    primaryClass = "physics.ins-det",
    reportNumber = "FERMILAB-PUB-19-259-CD-ND",
    doi = "10.1007/JHEP10(2019)052",
    journal = "JHEP",
    volume = "10",
    pages = "052",
    year = "2019"
}

@article{auger2014method,
  title={A method to suppress dielectric breakdowns in liquid argon ionization detectors for cathode to ground distances of several millimeters},
  author={Auger, Martin and Ereditato, Antonio and G{\"o}ldi, Damian and Janos, Stefan and Kreslo, Igor and L{\"u}thi, Matthias and von Rohr, C Rudolf and Strauss, Thomas and Tolba, Tamer and Weber, MS},
  journal={Journal of instrumentation},
  volume={9},
  number={07},
  pages={P07023},
  year={2014},
  publisher={IOP Publishing}
}

@article{haefliger2018comparison,
  title={Comparison of swarm and breakdown data in mixtures of nitrogen, carbon dioxide, argon and oxygen},
  author={Haefliger, Pascal and Franck, CM},
  journal={Journal of Physics D: Applied Physics},
  volume={52},
  number={2},
  pages={025204},
  year={2018},
  publisher={IOP Publishing}
}

@article{al2020electron,
  title={Electron transport in gaseous detectors with a Python-based Monte Carlo simulation code},
  author={Al Atoum, Bashar and Biagi, Steve F and Gonz{\'a}lez-D{\'i}az, Diego and Jones, Ben JP and McDonald, Austin D},
  journal={Computer Physics Communications},
  volume={254},
  pages={107357},
  year={2020},
  publisher={Elsevier}
}

@article{hunter1986electron,
  title={Electron transport measurements in methane using an improved pulsed Townsend technique},
  author={Hunter, SR and Carter, JG and Christophorou, LG},
  journal={Journal of applied physics},
  volume={60},
  number={1},
  pages={24--35},
  year={1986},
  publisher={American Institute of Physics}
}

@article{dahl2012obtaining,
  title={Obtaining precise electron swarm parameters from a pulsed Townsend setup},
  author={Dahl, Dominik A and Teich, Timm H and Franck, Christian M},
  journal={Journal of Physics D: Applied Physics},
  volume={45},
  number={48},
  pages={485201},
  year={2012},
  publisher={IOP Publishing}
}

@article{hornbeck1951microsecond,
  title={Microsecond transient currents in the pulsed Townsend discharge},
  author={Hornbeck, John A},
  journal={Physical Review},
  volume={83},
  number={2},
  pages={374},
  year={1951},
  publisher={APS}
}

@article{hernandez2003pulsed,
  title={Pulsed Townsend measurement of electron transport and ionization in SF6--N2 mixtures},
  author={Hernandez-Avila, JL and De Urquijo, J},
  journal={Journal of Physics D: Applied Physics},
  volume={36},
  number={12},
  pages={L51},
  year={2003},
  publisher={IOP Publishing}
}

@article{cooke1978nature,
  title={The nature and practice of gases as electrical insulators},
  author={Cooke, Chathan M and Cookson, Alan H},
  journal={IEEE Transactions on Electrical Insulation},
  number={4},
  pages={239--248},
  year={1978},
  publisher={IEEE}
}

@article{christophorou1976high,
  title={High voltage research (breakdown strengths of gaseous and liquid insulators)},
  author={Christophorou, LG and James, DR and Pai, RY and Pace, MO and Mathis, RA and Bouldin, DW},
  journal={Quarterly Report},
  year={1976}
}

@article{woodruff2020radio,
  title={Radio frequency and DC high voltage breakdown of high pressure helium, argon, and xenon},
  author={Woodruff, K and Baeza-Rubio, J and Huerta, D and Jones, BJP and McDonald, AD and Norman, L and Nygren, DR and Adams, C and {\'A}lvarez, V and Arazi, L and others},
  journal={Journal of Instrumentation},
  volume={15},
  number={04},
  pages={P04022},
  year={2020},
  publisher={IOP Publishing}
}

@article{rogers2018high,
  title={High voltage insulation and gas absorption of polymers in high pressure argon and xenon gases},
  author={Rogers, L and Clark, RA and Jones, BJP and McDonald, AD and Nygren, DR and Psihas, F and Adams, C and {\'A}lvarez, V and Arazi, L and Azevedo, CDR and others},
  journal={Journal of Instrumentation},
  volume={13},
  number={10},
  pages={P10002},
  year={2018},
  publisher={IOP Publishing}
}

@article{mcdonald2019electron,
  title={Electron drift and longitudinal diffusion in high pressure xenon-helium gas mixtures},
  author={McDonald, AD and Woodruff, K and Al Atoum, B and Gonz{\'a}lez-D{\'i}az, D and Jones, BJP and Adams, C and {\'A}lvarez, V and Arazi, L and Arnquist, IJ and Azevedo, CDR and others},
  journal={Journal of Instrumentation},
  volume={14},
  number={08},
  pages={P08009},
  year={2019},
  publisher={IOP Publishing}
}

@article{rogers2020mitigation,
  title={Mitigation of backgrounds from cosmogenic 137 Xe in xenon gas experiments using 3 He neutron capture},
  author={Rogers, L and Jones, BJP and Laing, A and Pingulkar, S and Smithers, B and Woodruff, K and Adams, C and {\'A}lvarez, V and Arazi, L and Arnquist, IJ and others},
  journal={Journal of Physics G: Nuclear and Particle Physics},
  volume={47},
  number={7},
  pages={075001},
  year={2020},
  publisher={IOP Publishing}
}

@article{felkai2018helium,
  title={Helium--Xenon mixtures to improve the topological signature in high pressure gas xenon TPCs},
  author={Felkai, R and Monrabal, F and Gonz{\'a}lez-D{\'i}az, Diego and Sorel, Michel and L{\'o}pez-March, N and G{\'o}mez-Cadenas, Juan Jos{\'e} and Adams, C and {\'A}lvarez, V and Arazi, L and Azevedo, CDR and others},
  journal={Nuclear Instruments and Methods in Physics Research Section A: Accelerators, Spectrometers, Detectors and Associated Equipment},
  volume={905},
  pages={82--90},
  year={2018},
  publisher={Elsevier}
}

@article{fernandes2020low,
  title={Low-diffusion Xe-He gas mixtures for rare-event detection: electroluminescence yield},
  author={Fernandes, AFM and Henriques, CAO and Mano, RDP and Gonz{\'a}lez-D{\'i}az, Diego and Azevedo, CD R and Silva, PAOC and G{\'o}mez-Cadenas, Juan Jos{\'e} and Freitas, Elisabete DC and Fernandes, LMP and Monteiro, CMB and others},
  journal={Journal of High Energy Physics},
  volume={2020},
  number={4},
  pages={1--18},
  year={2020},
  publisher={Springer}
}

@article{azevedo2016homeopathic,
  title={An homeopathic cure to pure Xenon large diffusion},
  author={Azevedo, CDR and Fernandes, LMP and Freitas, EDC and Gonzalez-Diaz, D and Monrabal, F and Monteiro, CMB and Dos Santos, JMF and Veloso, JFCA and Gomez-Cadenas, JJ},
  journal={Journal of Instrumentation},
  volume={11},
  number={02},
  pages={C02007},
  year={2016},
  publisher={IOP Publishing}
}

@article{gonzalez2015accurate,
  title={Accurate $\gamma$ and MeV-electron track reconstruction with an ultra-low diffusion Xenon/TMA TPC at 10 atm},
  author={Gonz{\'a}lez-D{\'i}az, Diego and {\'A}lvarez, V and Borges, FIG and Camargo, M and C{\'a}rcel, S and Cebri{\'a}n, Susana and Cervera, A and Conde, Carlos AN and Dafni, Theopisti and D{\'i}az, J and others},
  journal={Nuclear Instruments and Methods in Physics Research Section A: Accelerators, Spectrometers, Detectors and Associated Equipment},
  volume={804},
  pages={8--24},
  year={2015},
  publisher={Elsevier}
}

@article{dafni2009energy,
  title={Energy resolution of alpha particles in a microbulk Micromegas detector at high pressure argon and xenon mixtures},
  author={Dafni, Th and Ferrer-Ribas, E and Giomataris, I and Gorodetzky, Ph and Iguaz, F and Irastorza, IG and Salin, P and Tom{\'a}s, A},
  journal={Nuclear Instruments and Methods in Physics Research Section A: Accelerators, Spectrometers, Detectors and Associated Equipment},
  volume={608},
  number={2},
  pages={259--266},
  year={2009},
  publisher={Elsevier}
}

@article{cebrian2013micromegas,
  title={Micromegas-TPC operation at high pressure in xenon-trimethylamine mixtures},
  author={Cebrian, S and Dafni, T and Ferrer-Ribas, E and Giomataris, I and Gonzalez-Diaz, D and G{\'o}mez, H and Herrera, DC and Iguaz, FJ and Irastorza, IG and Luzon, G and others},
  journal={Journal of Instrumentation},
  volume={8},
  number={01},
  pages={P01012},
  year={2013},
  publisher={IOP Publishing}
}

@article{acciarri2014liquid,
  title={Liquid argon dielectric breakdown studies with the MicroBooNE purification system},
  author={Acciarri, R and Carls, B and James, C and Johnson, B and Jostlein, H and Lockwitz, S and Lundberg, B and Raaf, JL and Rameika, R and Rebel, B and others},
  journal={Journal of Instrumentation},
  volume={9},
  number={11},
  pages={P11001},
  year={2014},
  publisher={IOP Publishing}
}

@article{lockwitz2016study,
  title={A study of dielectric breakdown along insulators surrounding conductors in liquid argon},
  author={Lockwitz, Sarah and Jostlein, Hans},
  journal={Journal of Instrumentation},
  volume={11},
  number={03},
  pages={P03026},
  year={2016},
  publisher={IOP Publishing}
}

@article{blatter2014experimental,
  title={Experimental study of electric breakdowns in liquid argon at centimeter scale},
  author={Blatter, Alexander and Ereditato, Antonio and Hsu, C-C and Janos, S and Kreslo, I and Luethi, M and von Rohr, C Rudolf and Schenk, M and Strauss, T and Weber, MS and others},
  journal={Journal of instrumentation},
  volume={9},
  number={04},
  pages={P04006},
  year={2014},
  publisher={IOP Publishing}
}

@article{asaadi2014testing,
  title={Testing of high voltage surge protection devices for use in liquid argon TPC detectors},
  author={Asaadi, J and Conrad, JM and Gollapinni, S and Jones, BJP and Jostlein, H and St John, JM and Strauss, T and Wolbers, S and Zennamo, J},
  journal={Journal of Instrumentation},
  volume={9},
  number={09},
  pages={P09002},
  year={2014},
  publisher={IOP Publishing}
}

@article{cantini2017first,
  title={First test of a high voltage feedthrough for liquid Argon TPCs connected to a 300 kV power supply},
  author={Cantini, Cosimo and Gendotti, Adamo and Bueno, L Molina and Murphy, Sebastien and Radics, Balint and Regenfus, Christian and Rigaut, Yann-A and Rubbia, Andr{\'e} and Sergiampietri, Franco and Viant, Thierry and others},
  journal={Journal of Instrumentation},
  volume={12},
  number={03},
  pages={P03021},
  year={2017},
  publisher={IOP Publishing}
}

@article{tvrznikova2019direct,
  title={Direct comparison of high voltage breakdown measurements in liquid argon and liquid xenon},
  author={Tvrznikova, L and Bernard, EP and Kravitz, S and O'Sullivan, K and Richardson, G and Riffard, Q and Waldron, WL and Watson, J and McKinsey, DN},
  journal={Journal of Instrumentation},
  volume={14},
  number={12},
  pages={P12018},
  year={2019},
  publisher={IOP Publishing}
}

@article{von1957john,
  title={John Sealy Edward Townsend, 1868-1957},
  author={von Engel, Alfred},
  year={1957},
  pages="126-272",
  journal="Biographical Memoirs of Fellows of the Royal Society",
  publisher={The Royal Society London}
}

@inproceedings{martin2017pressurized,
  title={A pressurized argon gas TPC as DUNE near detector},
  author={Martin-Albo, J and DUNE Collaboration and others},
  booktitle={Journal of Physics: Conference Series},
  volume={888},
  number={1},
  pages={012154},
  year={2017},
  organization={IOP Publishing}
}

@article{Hamacher-Baumann:2020ogq,
    author = "Hamacher-Baumann, Philip and Lu, Xianguo and Mart{\'i}-Albo, Justo",
    title = "{Neutrino-hydrogen interactions with a high-pressure time projection chamber}",
    eprint = "2005.05252",
    archivePrefix = "arXiv",
    primaryClass = "physics.ins-det",
    doi = "10.1103/PhysRevD.102.033005",
    journal = "Phys. Rev. D",
    volume = "102",
    number = "3",
    pages = "033005",
    year = "2020"
}

@article{DUNE:2021tad,
    author = "Abed Abud, Adam and others",
    collaboration = "DUNE",
    title = "{Deep Underground Neutrino Experiment (DUNE) Near Detector Conceptual Design Report}",
    eprint = "2103.13910",
    archivePrefix = "arXiv",
    primaryClass = "physics.ins-det",
    reportNumber = "FERMILAB-PUB-21-067-E-LBNF-PPD-SCD-T",
    month = "3",
    year = "2021"
}

@article{NEXT:2019qbo,
    author = "Renner, J. and others",
    collaboration = "NEXT",
    title = "{Energy calibration of the NEXT-White detector with 1\% resolution near Q$\beta \beta$ of $^{136}$Xe}",
    eprint = "1905.13110",
    archivePrefix = "arXiv",
    primaryClass = "physics.ins-det",
    reportNumber = "FERMILAB-PUB-19-260-CD-ND",
    doi = "10.1007/JHEP10(2019)230",
    journal = "JHEP",
    volume = "10",
    pages = "230",
    year = "2019"
}

@article{MicroBooNE:2016pwy,
    author = "Acciarri, R. and others",
    collaboration = "MicroBooNE",
    title = "{Design and Construction of the MicroBooNE Detector}",
    eprint = "1612.05824",
    archivePrefix = "arXiv",
    primaryClass = "physics.ins-det",
    reportNumber = "FERMILAB-PUB-16-613-ND",
    doi = "10.1088/1748-0221/12/02/P02017",
    journal = "JINST",
    volume = "12",
    number = "02",
    pages = "P02017",
    year = "2017"
}

@article{DUNE:2015lol,
    author = "Acciarri, R. and others",
    collaboration = "DUNE",
    title = "{Long-Baseline Neutrino Facility (LBNF) and Deep Underground Neutrino Experiment (DUNE)}: {Conceptual Design Report, Volume 2: The Physics Program for DUNE at LBNF}",
    eprint = "1512.06148",
    archivePrefix = "arXiv",
    primaryClass = "physics.ins-det",
    reportNumber = "FERMILAB-DESIGN-2016-02",
    month = "12",
    year = "2015"
}

@article{ICARUS:2004wqc,
    author = "Amerio, S. and others",
    collaboration = "ICARUS",
    title = "{Design, construction and tests of the ICARUS T600 detector}",
    doi = "10.1016/j.nima.2004.02.044",
    journal = "Nucl. Instrum. Meth. A",
    volume = "527",
    pages = "329--410",
    year = "2004"
}

@article{LUX:2016ggv,
    author = "Akerib, D. S. and others",
    collaboration = "LUX",
    title = "{Results from a search for dark matter in the complete LUX exposure}",
    eprint = "1608.07648",
    archivePrefix = "arXiv",
    primaryClass = "astro-ph.CO",
    doi = "10.1103/PhysRevLett.118.021303",
    journal = "Phys. Rev. Lett.",
    volume = "118",
    number = "2",
    pages = "021303",
    year = "2017"
}

@article{EXO-200:2012pdt,
    author = "Auger, M. and others",
    collaboration = "EXO-200",
    title = "{Search for Neutrinoless Double-Beta Decay in $^{136}$Xe with EXO-200}",
    eprint = "1205.5608",
    archivePrefix = "arXiv",
    primaryClass = "hep-ex",
    doi = "10.1103/PhysRevLett.109.032505",
    journal = "Phys. Rev. Lett.",
    volume = "109",
    pages = "032505",
    year = "2012"
}

@article{Aprile:2012zx,
    author = "Aprile, Elena",
    editor = "Cline, David",
    collaboration = "XENON1T",
    title = "{The XENON1T Dark Matter Search Experiment}",
    eprint = "1206.6288",
    archivePrefix = "arXiv",
    primaryClass = "astro-ph.IM",
    doi = "10.1007/978-94-007-7241-0-14",
    journal = "Springer Proc. Phys.",
    volume = "148",
    pages = "93--96",
    year = "2013"
}

@article{NEXT:2012zwy,
    author = "Alvarez, V. and others",
    collaboration = "NEXT",
    title = "{NEXT-100 Technical Design Report (TDR): Executive Summary}",
    eprint = "1202.0721",
    archivePrefix = "arXiv",
    primaryClass = "physics.ins-det",
    doi = "10.1088/1748-0221/7/06/T06001",
    journal = "JINST",
    volume = "7",
    pages = "T06001",
    year = "2012"
}

@article{uhm2000breakdown,
  title={Breakdown properties of high-pressure electrical discharge},
  author={Uhm, Han S and Choi, Eun H and Cho, Guangsup},
  journal={Physics of Plasmas},
  volume={7},
  number={6},
  pages={2744--2746},
  year={2000},
  publisher={American Institute of Physics}
}

@book{raizeryurip,
    title={Gas discharge physics}, 
    publisher={Springer}, 
    author={Razier, Yuri P.},
    year={1997}
}

@article{csahin2016systematic,
  title={Systematic gas gain measurements and Penning energy transfer rates in Ne- CO2 mixtures},
  author={{\c{S}}ahin, {\"O} and Kowalski, TZ and Veenhof, R},
  journal={Journal of Instrumentation},
  volume={11},
  number={01},
  pages={P01003},
  year={2016},
  publisher={IOP Publishing}
}

\end{document}